\documentclass{llncs}
\usepackage{makeidx}  
\usepackage{epsfig}  
\begin{document}

\title{A generating function method for the average-case analysis of DPLL}
\author{R\'emi Monasson}
\institute{CNRS-Laboratoire de Physique Th{\'e}orique, ENS, Paris, France
\\{\bf monasson@lpt.ens.fr}}
\maketitle

\begin{abstract}
A method to calculate the average size of
Davis-Putnam-Loveland-Logemann (DPLL) search trees for random
computational problems is introduced, and applied to the
satisfiability of random CNF formulas (SAT) and the coloring of random
graph (COL) problems. We establish recursion relations for the generating
functions of the average numbers of (variable or color) assignments at a given
height in the search tree, which allow us to derive the asymptotics of 
the expected DPLL tree size, $2^{\, N \omega + o(N)}$, where $N$ is the
instance size. $\omega$ is calculated as a function of the
input distribution parameters (ratio of clauses per variable for
SAT, average vertex degree for COL), and the branching heuristics.  
\end{abstract}

\baselineskip=16pt

\section{Introduction and main results.}

Many efforts have been devoted to the study 
of the performances of the Davis-Putnam-Loveland-Logemann (DPLL) 
procedure \cite{Karp}, and more generally, 
resolution proof complexity for combinatorial problems 
with randomly generated instances. Two examples
 are random $k$-Satisfiability ($k$-SAT), where an 
instance ${\cal F}$ is a uniformly and randomly chosen set of $M=\alpha\, N$
disjunctions of $k$ literals  built from $N$ Boolean variables and 
their negations (with no repetition and no complementary literals), and 
random graph $k$-Coloring ($k$-COL), where an instance ${\cal F}$ is 
an Erd\H{o}s-R\'enyi random graph from $G(N,p=c/N)$ {\em i.e.} with 
average vertex degree  $c$.

Originally, efforts were concentrated on the random width 
distribution for $k$-SAT, where each literal appear with a fixed probability.  
Franco, Purdom and collaborators  showed that 
simplified versions of DPLL had polynomial average-case complexity in 
this case, see \cite{purd2,Fra2} for reviews. It was then recognized that
the fixed clause length ensemble might provide harder instances
for DPLL \cite{fra0}. Chv\'atal and
Szemer\'edi indeed showed that DPLL proof size is w.h.p. exponentially
large (in $N$ at fixed ratio $\alpha$) for an unsatisfiable instance
\cite{Chv}.  Later on, Beame {\em et al.}  \cite{Bea} showed that the
proof size was w.h.p. bounded from above by $2^{c \, N/\alpha}$ (for
some constant $c$), a decreasing function of $\alpha$. As for the
satisfiable case, Frieze and Suen showed that backtracking is
irrelevant at small enough ratios $\alpha$ ($\le 3.003$ with the
Generalized Unit Clause heuristic, to be defined below) \cite{Fri},
allowing DPLL to find satisfying assignment in polynomial (linear)
time. Achlioptas, Beame and Molloy proved that, conversely, at ratios
smaller than the generally accepted satisfiability
threshold, DPLL takes w.h.p. exponential time to find a satisfying assignment
\cite{Achl3}.  Altogether these results provide explanations for the
`easy-hard-easy' (or, more precisely, `easy-hard-less hard') pattern
of complexity experimentally observed when running DPLL on random
3-SAT instances \cite{Mit}. 

A precise calculation of the average size of the search space explored by 
DPLL (and \#DPLL, a version of the procedure solving the enumeration 
problems \#SAT and \#COL) as a function of the parameters
$N$ and $\alpha$ or $c$ is difficult due to the 
statistical correlations between branches in the search tree resulting 
from backtracking. Heuristic derivations  were
nevertheless proposed by Cocco and Monasson based on a `dynamic annealing' 
assumption  \cite{Coc,Coc2,Eindor}. Hereafter, using the linearity of
expectation, we show that `dynamic annealing' turns not to be an 
assumption at all when the expected tree size is concerned.

We first illustrate the approach, based on the use of recurrence relations 
for the generating functions of the number of nodes at a given height in 
the tree, on the random $k$-SAT problem and
the simple Unit Clause (UC) branching heuristic where unset variables 
are chosen uniformly
at random and assigned to True or False uniformly at random 
\cite{fra2,Fra}. Consider the following counting
algorithm   
\vskip .3cm \noindent
{\small Procedure \#DPLL-UC[${\cal F}$,A,S]

\noindent Call ${\cal F}_A$ what is left from instance ${\cal F}$ given
partial variable assignment $A$;

1. If ${\cal F}_A$ is empty, $S\to S+ 2^{N-|A|}$, Return; 
{\em (Solution Leaf)}

2. If there is an empty clause in ${\cal F}_A$, Return; 
{\em (Contradiction Leaf)}

3. If there is no empty clause in ${\cal F}_A$,
let $\Gamma _1=\{$1-clauses$\ \in {\cal F}_A\}$,

\hskip .6cm if $\Gamma _1 \ne \emptyset$, pick any 1-clause, say,  $\ell$,
and call DPLL[${\cal F}$,A$\cup \ell$]; {\em (unit-propagation)}

\hskip .6cm  if $\Gamma_1=\emptyset$, pick up  an unset
literal uniformly at random, say, $\ell$, and call

\hskip 2.1cm DPLL[${\cal F}$,A$\cup \ell$], then
DPLL[${\cal F}$,A$\cup \bar \ell$] ; {\em (variable splitting)}

\noindent End;}

\noindent
\#DPLL-UC, called with $A=\emptyset$ and $S=0$, returns the number $S$
of solutions of the instance ${\cal F}$; the history of the search 
can be summarized as a search tree with leaves marked with solution or 
contradiction labels.  As the instance to be treated and the sequence
of operations done by \#DPLL-UC are stochastic, so are
the numbers $L_S$ and $L_C$ of solution and contradiction leaves respectively.

\begin{theorem}
{Let $k\ge 3$ and 
$\displaystyle{
\Omega (t,\alpha,k) =  t + \alpha \; \log_2 \big(1
- \frac {k}{2^k} t^{k-1} +\frac{k-1}{2^k} {t^k} \big)}$.
The expectations of the 
numbers of solution and contradiction leaves in the 
\#DPLL-UC search tree of random $k$-SAT instances with $N$ variables
and $\alpha N$ clauses are, respectively,
$\displaystyle{
L_S(N,\alpha,k)=2^{N \omega _S(\alpha,k) +o(N)}}$ with 
$\omega _S(\alpha,k)= \Omega (1,\alpha,k)$ and
$\displaystyle{ 
L_C(N,\alpha,k)=2^{N \omega _C (\alpha ,k) +o(N)}  \ \ with \ \
\omega _C(\alpha,k)=\max _{t\in [0;1]}\Omega (t,\alpha,k)
}$.}
\end{theorem}
\noindent
An immediate consequence of Theorem 1 is that 
the expectation value of the total number of leaves, $L_S+L_C$, is
$2^{N \omega _C (\alpha ,k) +o(N)}$. This result was 
first found by M\'ejean, Morel and Reynaud in the particular case $k= 3$ 
and for ratios 
$\alpha > 1$ \cite{mejean}. Our approach not only provides a much
shorter proof, but can also be easily extended to other
problems and more sophisticated heuristics, see Theorems 2 and 3 below.
In addition, Theorem 1 provides us with some information about the expected
search tree size of the decision procedure DPLL-UC, 
corresponding  to \#DPLL-UC with Line 1 replaced with:
{\small If ${\cal F}_A$ is empty, output Satisfiable; Halt}.
\begin{corollary}
{Let $\alpha > \alpha _u(k)$, the root
of $\omega _C (\alpha,k) = 2+ \alpha \log _2(1-2^{-k})$ e.g. 
$\alpha _u (3)= 10.1286...$. The average size of 
DPLL-UC search trees for random $k$-SAT instances 
with $N$ variables and $\alpha N$ clauses equals $2^{N \omega _C(\alpha,k) 
+o(N)}$.}
\end{corollary}

\noindent
Functions $\omega_S,\omega_C$  are shown in Figure 1 in the $k=3$ case. They 
coincide and are equal to $1 - \alpha
\log_2 (8/7)$ for $\alpha < \alpha ^* = 4.56429...$, while 
$\omega_C > \omega_S$ for $\alpha > \alpha ^*$. In other words, for
$\alpha >\alpha^*$, most leaves in \#DPLL-UC trees are contradiction
leaves, while for $\alpha < \alpha ^*$, both contradiction and
solution leaf numbers are (to exponential order in $N$) of the same
order. As for DPLL-UC trees, notice that 
$\displaystyle{
\omega _C(\alpha,k) \asymp \frac{2 \ln 2}{3 \,\alpha} =
\frac{0.46209...}\alpha
}$.
This behaviour agrees with 
Beame et al.'s result ($\Theta (1/\alpha)$)  for the average resolution 
proof complexity of unsatisfiable instances \cite{Bea}. 
Corollary 1 shows that the expected DPLL tree size can be estimated
for a whole range of $\alpha$; we conjecture that the above expression 
holds for ratios smaller than $\alpha _u$ {\em i.e.}
down to $\alpha ^*$ roughly. For generic $k\ge 3$,
we have $\displaystyle{
\omega _C(\alpha,k) \asymp \frac{k-2}{k-1} \bigg( 
\frac{2^k \ln 2}{k (k-1) \, \alpha} \bigg) ^{1/(k-2)}}$; the decrease
of $\omega_C$ with $\alpha$ is therefore slower and slower as $k$ increases.

\begin{figure}[h]
\begin{center}
\epsfig{file=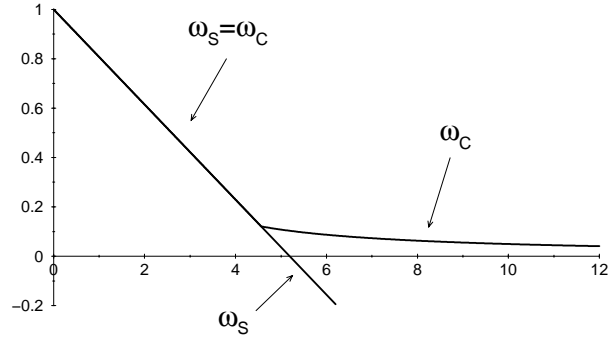,angle=-90,width=8cm}
\end{center}
\caption{Logarithms of the average numbers of solution and
contradiction leaves, respectively $\omega _S$ and $\omega_C$, in
\#DPLL-UC search trees versus ratio $\alpha$ of
clauses per variable for random 3-SAT. 
Notice that $\omega_C$ coincides with the
logarithm of the expected size of \#DPLL-UC at all ratios $\alpha$, and 
with the one of DPLL-UC search trees for $\alpha \ge 10.1286...$.}
\label{figomega}
\end{figure}

So far, no expression for $\omega$ has been obtained for more sophisticated 
heuristics than UC. We consider 
the Generalized Unit Clause (GUC) heuristic \cite{Fra,Achl}
where the shortest clauses are preferentially satisfied. The associated
decision procedure, DPLL-GUC, corresponds to DPLL-UC with Line 3 replaced with:
{\small Pick a clause uniformly at random  among the shortest clauses, 
and a literal, say, $\ell$, in the clause; call 
DPLL[${\cal F}$,A$\cup \ell$], then DPLL[${\cal F}$,A$\cup \bar \ell$].}
\begin{theorem}
Define 
$m(x_2) = \frac 12 \big(1 + \sqrt{1+4x_2}) -2 x_2$,
$y_3(y_2)$ the solution of the ordinary differential equation 
${dy_3}/{dy_2} = 3(1+y_2-2 \;y_3)/(2m(y_2))$
such that $y_3(1)=1$, and
\begin{equation} \nonumber \label{refiuy}
\omega ^g (\alpha) = \max _{\frac 34 < y_2 \le 1} \bigg[ 
\int _{y_2}^1 \frac {dz}{m(z)} \log _2 \big(2z +m(z)\big)\ 
\exp \left( - \int _{z}^1 \frac {dw}{m(w)}\right)
+ \alpha \log _2 y_3 (y_2)   \bigg] \quad .
\end{equation}
Let $\alpha > \alpha ^g _u  = 10.2183...$, the root 
of $\omega ^g (\alpha) + \alpha \log _2(8/7) =2$. 
The expected size of DPLL-GUC search tree for random 3-SAT instances 
with $N$ variables and $\alpha N$ clauses is $2^{N\,\omega ^g (\alpha) +
o(N)}$.
\end{theorem}

\noindent
Notice that, at large $\alpha$,
$\displaystyle{
\omega ^g  (\alpha ) \asymp 
\frac{3+\sqrt{5}}{6\,\ln 2}\big[ \ln \big( \frac{1+\sqrt{5}}{2} \big)
\big]^2 \frac 1\alpha = \frac {0.29154...}\alpha 
}$
in agreement with the $1/\alpha$ scaling established in \cite{Bea}. 
Furthermore, the multiplicative factor is smaller than the one for
UC, showing that DPLL-GUC is more efficient than DPLL-UC in
proving unsatisfiability.
\vskip .3cm
A third application is the analysis of the counterpart of GUC for the
random 3-COL problem. The version of DPLL we have analyzed operates as
follows \cite{mol}. 
Initially, each vertex is assigned a list of 3 available colors. 
In the course of the procedure, a vertex, say, $v$,  with
the smallest number of available colors, say, $j$, is chosen at random 
and uniformly. DPLL-GUC then removes $v$, and successively branches to the 
$j$ color
assignments corresponding to removal of one of the $j$ colors of $v$ from the
lists of the neighbors of $v$. The procedure backtracks when a vertex
with no color left is created ({contradiction}), or no vertex
is left (a {proper coloring} is found). 
\begin{theorem}
Define 
$\displaystyle{
\omega ^h (c) = \max _{0 < t < 1} \big[ 
\frac c6 t^2 - \frac c3 t - (1-t) \ln 2 + \ln \big( 3 -e^{-2 c\,t / 3}  
\big) \big]
}$.

\noindent
Let $c > c ^h _u  = 13.1538...$, the root 
of $\omega ^h (c) + \frac c6 = 2 \ln 3$. 
The expected size of DPLL-GUC search tree for deciding 3-COL on random 
graphs from $G(N,c/N)$ with $N$ vertices is $e^{\,N\,\omega ^h (c) +
o(N)}$.
\end{theorem}

\noindent
Asymptotically, $\displaystyle{
\omega ^h  (c ) \asymp 
\frac{3\, \ln 2}{2\, c^2}= \frac {1.0397...}{c^2} 
}$
in agreement with Beame et al.'s scaling ($\Theta (1/c^2)$) \cite{Cris}. 
An extension of Theorem 3 to higher values of the number $k$ of 
colors gives 
$\displaystyle{ \omega^h(c,k) \asymp 
\frac{k(k-2)}{k-1} \; \big[ \frac{2\, \ln 2}{k-1}
\big] ^{1/(k-2)} c^{-(k-1)/(k-2)}}$. This result is compatible with
the bounds derived in \cite{Cris}, and  suggests that the  
$\Theta( c ^{-(k-1)/(k-2)})$ dependence 
could hold w.h.p. (and not only 
in expectation).


\section{Recurrence equation for \#DPLL-UC search tree}

Let ${\cal F}$ be an instance of the 3-SAT problem defined over a set of
$N$ Boolean variables $X$. A partial assignment $A$ 
of length $T (\le N)$ is the specification of the truth values of $T$ 
variables in $X$. We denote by ${\cal F}_A$ the residual instance 
given $A$. A clause $c\in {\cal F}_A$ is said to be a $\ell$-clause
with $\ell \in\{0,1,2,3\}$ if the number of false literals 
in $c$ is equal to $3-\ell$.  We denote by $C _\ell({\cal F}_A)$ 
the number of $\ell$-clauses in ${\cal F}_A$. 
The instance ${\cal F}$ is said to be satisfied under $A$ if 
$C_\ell ({\cal F}_A)=0$ for $\ell=0,1,2,3$, unsatisfied (or violated)
under $A$ if $C_0({\cal F}_A) \ge 1$,  undetermined under $A$ otherwise. 
The clause vector of an undetermined or satisfied residual instance
${\cal F}_A$ is the three-dimensional vector $\vec C$ with components
$C_1({\cal F}_A),C_2({\cal F}_A),C_3({\cal F}_A)$. 
The search tree associated to an instance ${\cal F}$
and a run of \#DPLL is the tree whose nodes carry the residual assignments 
$A$ considered in the course of the search. 
The height $T$ of a node is the length of the attached assignment.

\vskip .3cm

It was shown by Chao and Franco \cite{fra2,Fra}
that, during the first descent in the search tree {\em i.e.} prior to any 
backtracking, the distribution of residual instances remains uniformly
random conditioned on the numbers of $\ell$-clauses. This statement remains
correct for heuristics more sophisticated than UC {\em e.g.}
GUC, SC$_1$ \cite{Fra,Achl}, and was recently extended to splitting 
heuristics based on variable occurrences by Kaporis, Kirousis and 
Lalas \cite{Kap}. Clearly, in this context, uniformity is lost after 
backtracking enters into play (with the exception of Suen and Frieze's 
analysis of a limited version of backtracking \cite{Fri}). 
Though this limitation appears to forbid (and has forbidden so far) 
the extension of average-case studies of backtrack-free DPLL to full DPLL 
with backtracking, we point out here that it is not as severe as it
looks. Indeed, let us forget about how \#DPLL or DPLL search tree is built and
consider its final state. We refer to a
branch (of the search tree) as the shortest path from the root node
(empty assignment) to a leaf. The two key remarks underlying the
present work can be informally stated as follows.
First, the expected size of a \#DPLL search tree can be calculated from the
knolwedge of the statistical distribution of (residual instances on) 
a single branch; no characterization of the 
correlations between distinct branches in the tree is necessary.
Secondly, the statistical distribution of (residual instances on) a single 
branch is simple since, along a branch, uniformity is 
preserved (as in the absence of backtracking).
More precisely,
\begin{lemma}
[from Chao \& Franco \cite{fra2}] 
{ Let ${\cal F}_A$ be a residual instance attached
to a node $A$ at height $T$ in a \#DPLL-UC search tree
produced from an instance ${\cal F}$ drawn from the random 3-SAT distribution. 
Then the set of $\ell$-clauses in ${\cal F}_A$ is
uniformly random conditioned on its size $C_\ell({\cal F}_A)$
and the number $N-T$ of unassigned variables for each $\ell \in\{0,1,2,3\}$.}
\end{lemma}
\begin{proof}
 the above Lemma is an immediate application 
of Lemma 3 in Achlioptas' Card Game framework which establishes
uniformity for algorithms ({\em a}) `pointing to a particular card (clause)', 
or ({\em b}) 'naming a variable that has not yet been assigned a value' 
(Section 2.1 in Ref. \cite{Achl}). 
The operation of \#DPLL-UC along a branch precisely amounts to 
these two operations: unit-propagation relies on action ({\em a}), and
variable splitting on ({\em b}). \qed
\end{proof}
Lemma 1 does not address the question of uniformity among 
different branches. Residual instances attached to two (or more)
nodes on distinct branches in the search tree are correlated.
However, these correlations can be
safely ignored in calculating the average number of residual instances, 
in much the same way as the average
value of the sum of correlated random variables is simply the sum of
their average values. 
\begin{proposition} 
{Let $L ( \vec C , T)$ be the
expectation of the number of undetermined residual instances with clause 
vector $\vec C$ at height $T$ in \#DPLL-UC search tree, and  
$\displaystyle{
G (x_1,x_2,x_3;T\,) =\sum_{\vec C}\; x_1 ^{\, C_1}\; x_2 ^{\, C_2}\; 
x_3 ^{\, C_3}\ L(\,{\vec C}\,,T\,)
}$
its generating function. Then, for $0\le T <N$,
\vskip -.4cm
\begin{eqnarray}
\label{eqev}
G(x_1,x_2,x_3;T+1\,)&=& \frac 1{f_1}\; G \big( f_1,f_2,f_3; 
T\,\big) + \bigg( 2 - \frac 1{f_1} \bigg) \; G\big( 0, f_2,f_3; T\, \big) 
\nonumber \\
&-& 2 \;  G(0,0,0;T) 
\end{eqnarray}
where $f_1,f_2,f_3$ stand for the functions 
$f_1 ^{(T)} (x_1)=x_1+ \frac 12 \mu (1 -2 x_1)$,
$f_2 ^{(T)}  (x_1,x_2)=x_2+ \mu ( x_1+1 -2x_2)$,
$f_3  ^{(T)}(x_2,x_3)=x_3+\frac32 \mu ( x_2+1 -2x_3)$,
and $\mu =1/(N-T)$. The generating function
$G$ is entirely defined from recurrence relation (\ref{eqev})
and the initial condition $G(x_1,x_2,x_3 ; 0) =\big( x_3 \big)^{\alpha N}$.}
\end{proposition}
\begin{proof} Let $\delta _n$  denote the Kronecker function 
($\delta _n=1$ if $n=0$, $\delta _n=0$ otherwise),
$B _n ^{m,q}={m \choose n} q^n(1-q)^{m-n}$ the 
binomial distribution. Let $A$ be a node
at height $T$, and ${\cal F}_A$ the attached residual instance.
Call $\vec C$ the clause vector of ${\cal F}_A$. Assume first that 
$C_1\ge 1$. Pick up one 1-clause, say, $\ell$. Call $z_j$
the number of $j$-clauses that contain $\bar \ell$ or $\ell$ 
(for $j=1,2,3$). From Lemma 1, the $z_{j}$'s are 
binomial variables with parameter $j/(N-T)$ among $C_j-\delta _{j-1}$
(the 1-clause that is satisfied through unit-propagation is removed). 
Among the $z_j$ clauses, $w_{j-1}$ contained $\bar \ell$ and are
reduced to $(j-1)$-clauses, while the remaining $z_j-w_{j-1}$ contained
$\ell$ and are satisfied and removed. From Lemma 1 again, $w_{j-1}$ is
a binomial variable with parameter $1/2$ among $z_j$.
The probability that the instance produced has no
empty clause ($w_0=0$) is $B_0^{z_1,\frac 12}=2^{-z_1}$. Thus, setting 
$\mu = \frac 1{N-T}$,
\begin{eqnarray}
\label{bbra2}
 M _{P} && [\vec C',\vec C; T] =
\sum _{z_3=0} ^{C_3} B_{z_3}^{ C_3,3\mu} \sum _{w_2=0}^{z_3} 
B_{w_2}^{z_3,\frac 12} \sum_{z_2=0}^{C_2} B_{z_2}^{ C_2,2\mu} 
\sum _{w_1=0}^{z_2} B_{w_1}^{z_2,\frac 12} \nonumber \\ && 
\times\ \sum _{z_1=0}^{C_1-1} B_{z_1}^{C_1-1,\mu} 
\frac 1{2^{z_1}} \, \delta _{C'_3 - (C_3 - z_3)}
\delta _{C'_2- (C_2 -z_2+w_2)} \delta _{C' _1- (C_1-1 -z_1+w_1)} \nonumber
\end{eqnarray}
expresses the probability that a residual instance at height $T$ with
clause vector $\vec C$ gives rise to a (non-violated) residual instance 
with clause vector $\vec C'$  at height $T+1$ through unit-propagation. 
Assume now $C_1=0$. Then, a
yet unset variable is chosen and set to True or False uniformly at 
random. The calculation of the new vector $\vec C'$ is identical
to the unit-propagation case above, except that: $z_1=w_0=0$ (absence
of 1-clauses), and two nodes are produced (instead of one).
Hence,
\begin{eqnarray}
M _{UC}  [\vec C',\vec C; T] &=& 2\sum _{z_3=0} ^{C_3} B_{z_3}^{ C_3,3\mu}
\sum _{w_2=0}^{z_3} B_{w_2} ^{z_3,\frac 12} 
\sum_{z_2=0}^{C_2} B_{z_2}^{ C_2,2\mu} 
\sum _{w_1=0}^{z_2} B_{w_1}^{z_2,\frac 12} \nonumber \\
&&\times\ \delta _{C'_3 - (C_3 - z_3)} \delta _{C'_2- (C_2 -z_2+w_2)}
\delta _{C' _1- w_1 }  \nonumber 
\end{eqnarray}
expresses the expected number of residual instances at height $T+1$ 
and with clause vector $\vec C'$ produced from a residual instance at 
height $T$ and with clause vector $\vec C$ through UC  branching. 

Now, consider all the nodes $A_i$ at height $T$, with $i=1,
\ldots, {\cal L}$. Let  $o_i$ be the operation done by \#DPLL-UC
on $A_i$. $o_i$ represents either unit-propagation (literal $\ell_i$
set to True) or variable splitting (literals $\ell _i$ set to T and F
on the descendent nodes respectively).    
Denoting by $\mathbf{E}_{Y}(X)$ the
expectation value of a quantity $X$ over variable $Y$,  
$\displaystyle{
L(\vec C';T+1) = \mathbf{E}_{{\cal L}, \{A_i, o_i\}} \left(
\sum _{i=1}^{\cal L} {\cal M} [\vec C' ; A_i,o_i ] \right)
}$
where ${\cal M}$ is the number (0, 1 or 2) of
residual instances with clause vector $\vec C'$ produced from $A_i$ 
after \#DPLL-UC has carried out operation $o_i$.
Using the linearity of expectation, 
$\displaystyle{ 
L(\vec C';T+1) = \mathbf{E}_{{\cal L}} \left(\sum _{i=1}^{\cal L} 
\mathbf{E}_{\{A_i, o_i\}} \big( {\cal M} [\vec C' ; A_i,o_i ] \big)\right)
= \mathbf{E}_{{\cal L}} \left( \sum _{i=1}^{\cal L} 
M[\vec C',\vec C_i; T]
\right)}
$

\noindent
where $\vec C_i$ is the clause vector of the residual
instance attached to $A_i$, and
$M[\vec C',\vec C; T] = \big( 1-\delta_{C_1} \big)  
 \, M _{P} [\vec C',\vec C; T] 
+ \delta _{C_1}\,  M_{UC} [\vec C',\vec C; T]$. 
Gathering assignments with identical clause vectors gives the 
reccurence relation
$L ( \vec C' , T+1) =$
$\displaystyle{\sum_{\vec C}  M [\vec C',\vec C; T] 
 \  L ( \vec C , T)}$.
Recurrence relation (\ref{eqev}) for the generating function is an immediate
consequence.  The initial condition over $G$ stems from the fact that
the instance is originally drawn from the random 3-SAT distribution,  
$L(\vec C;0)=\delta _{C_1}\, \delta _{C_2}\, \delta_{C_3-\alpha N}$.\qed
\end{proof}

\section{Asymptotic analysis and application to DPLL-UC}

The asymptotic analysis of $G$  relies on the 
following technical lemma:
\begin{lemma} 
{Let $\gamma (x_2,x_3,t)  =  (1-t)^3 x_3 
+ \frac {3t}2 (1-t)^2 x_2+ \frac t8 (12-3t-2t^2)$, with
$t\in]0;1[$ and $x_2,x_3>0$. Define 
$\displaystyle{S_0(T) \equiv \sum _{H=0}^{T} 2^{T-H} \, G(0,0,0;H)}$.
Then, in the large $N$ limit, 
$\displaystyle{
S_0([tN]) \le 2^{N (t + \alpha \log _2 \gamma(0,0,t)) +
o(N)}}$ and $G \big(\frac 12,x _2,x_3;[tN]\big) =  
\displaystyle{2^{N(t + \alpha \log _2 \gamma(x_2,x_3,t)) + o(N)}}$.}
\end{lemma}
Due to space limitations, we give here only some elements of the
proof.
The first step in the proof is inspired 
by Knuth's kernel method \cite{knu}: when $x_1=\frac 12$, $f_1=\frac 12$ and 
recurrence relation (\ref{eqev}) simplifies and is easier to handle. Iterating
this equation then allows us to relate the value of $G$ at height $T$ and
coordinates $(\frac 12, x_2,x_3)$ to the (known) value of $G$ at height 0
and coordinates $(\frac 12,y_2,y_3)$ which are functions of $x_2,x_3, T,N$,
and $\alpha$. The function $\gamma$ is the value of $y_3$ when $T,N$
are sent to infinity at fixed ratio $t$. The asymptotic statement
about $S_0(T)$ comes from the previous result and the fact that the
dominant terms in the sum  defining $S_0$ are the ones 
with $H$ close to $T$. 

\begin{proposition}
{Let $L_C(N,T,\alpha)$ be the expected number of contradiction leaves
of height $T$ in the \#DPLL-UC resolution tree of random 3-SAT instances
with $N$ variables and $\alpha N$ clauses, and $\epsilon >0$.
Then, for $t\in [\epsilon; 1-\epsilon]$ and $\alpha >0$,
$\displaystyle{
\Omega (t,\alpha,3) \le 
\frac 1N \log _2 L_C(N,[tN],\alpha) +o(1) \le \max _{h\in[\epsilon, ;t]}  
\Omega (h,\alpha,3) 
}$
where $\Omega$ is defined in Theorem 1.
}
\end{proposition}
\noindent
Observe that a contradiction may appear with a positive (and 
non--exponentially small in $N$) probability as soon as two 1-clauses are 
present. These 1-clauses will be present as a result of 2-clause 
reduction when the residual instances include a large number ($\Theta (N)$) 
of 2-clauses. As this is the case for a finite fraction of residual
instances, $G(1,1,1;T)$ is not exponentially larger than $L_C(T)$. Use of
the monotonicity of $G$ with respect to $x_1$ and Lemma 2 gives the announced
lower bound (recognize that $\Omega (t,\alpha,3)=t + \alpha \log _2
\gamma (1,1;t)$). To derive the upper bound, remark that 
contradictions leaves cannot be more numerous
than the number of branches created through splittings; hence $L_C(T)$ 
is bounded from above by the number of splittings at 
smaller heights $H$, that is, $\displaystyle{\sum _{H < T}} G(0,1,1;H)$.
Once more, we use the monotonicity of $G$ with respect to $x_1$ and
Lemma 2 to obtain the upper bound.
The complete proof will be given in the full version.

\begin{proof} {\em (Theorem 1)}
By definition, a solution leaf is a node in the search tree  
where no clauses are left; the average number $L_S$ of solution leaves 
is thus given by
$\displaystyle{
L_S = \sum _{H=0}^N L(0,0,0;H) = \sum _{H=0}^N G(\vec 0;H)
}$. A straightforward albeit useful upper bound on $L_S$ is obtained from
$L_S  \le  S_0(N)$.
By definition of the algorithm \#DPLL, $S_0 (N)$ is
the average number of solutions of an instance with $\alpha N$ clauses
over $N$ variables drawn from the random 3-SAT distribution,
$S_0 (N) = 2^N \,(7/8) ^{\alpha N}$ \cite{fra0}.
This upper bound is indeed tight (to within terms that are 
subexponential in $N$), 
as most solution leaves have heights equal, or close to $N$. To show this, 
consider $\epsilon >0$, and write 
$$
L_S \ge \sum _{H=N(1-\epsilon)}^N G(\vec 0;H) 
\ge 2^{-N \epsilon} \;  \sum _{H=N(1-\epsilon)}^N 2^{N-H} G(\vec 0;H)
=  2^{-N \epsilon} \;   S_0(N) \; \big[ 1 - A\big]  
$$
with $A = 2^{N\epsilon} S_0(N(1-\epsilon))/S_0(N)$. From 
Lemma 2, $A\le (\kappa +o(1))^{\alpha N}$ with
$\displaystyle{
\kappa  = \frac{\gamma(0,0,1-\epsilon)}{7/8} = 1 - \frac 97 \, \epsilon^2
+ \frac 27 \, \epsilon ^3 <1
}$
for small enough $\epsilon$ (but $\Theta(1)$ with respect to $N$).
We conclude that $A$ is exponential small in $N$, and 
$
-\epsilon + 1 - \alpha \log_2 \frac 87 + o(1) 
\le \frac 1N \log_2 L_S \le  1 - \alpha \log_2 \frac 87 
$.
Choosing arbitrarily small $\epsilon$ allows us to
establish the statement about the asymptotic behaviour of
$L_S$ in Theorem~1.

Proposition 2, with arbitrarily small $\epsilon$, immediately
leads to Theorem 1 for $k=3$, 
for the average number of contradiction leaves, $L_C$, equals
the sum over all heights $T=tN$ (with $0\le t\le 1$) of $L_C(N,T,\alpha)$,
and the sum is bounded from below by its largest term and,
from above, by $N$ times this largest term. 
The statement on the number of leaves following Theorem 1
 comes from the observation that the expected 
total number of leaves is $L_S+L_C$, and $\displaystyle{\omega _S 
(\alpha,3) = \Omega (1,\alpha,3) \le \max _{t \in[0;1]} \Omega
(t,\alpha,3) = \omega _C(\alpha,3 )}$.
\qed
\end{proof}

\begin{proof} ({\em Corollary 1}) Let $P_{sat}$ be the probability
that a random 3-SAT instance with $N$ variables and $\alpha N$ clauses
is satisfiable. Define
$\#L_{sat}$ and $\#L_{unsat}$ (respectively, $L_{sat}$ and
$L_{unsat}$) the expected numbers of leaves in \#DPLL-UC (resp. DPLL-UC) 
search trees for satisfiable and unsatisfiable instances respectively. 
All these quantities depend on $\alpha$ and $N$. 
As the operations of \#DPLL and DPLL coincide for unsatifiable instances, 
we have $\#L_{unsat} =L_{unsat}$. Conversely, $\#L_{sat} \ge L_{sat}$ since
DPLL halts after having encountered the first solution leaf.
Therefore, the difference between the average sizes \#L and L 
of \#DPLL-UC and DPLL-UC search trees satisfies
$0 \le \#L - L =  P_{sat} \; (\#L_{sat} - L_{sat}) \le  P_{sat} \; \#L_{sat}$.
Hence, $1 -  P_{sat} \; \#L_{sat}/\#L \le L/\#L \le 1$.
Using $\#L_{sat}\le 2^N$, $P_{sat} \le 2^N (7/8)^{\alpha N}$ from the
first moment theorem and the asymptotic scaling for $\#L$ given in
Theorem 1, we see that the left hand side of the previous inequality tends to 1
when $N\to \infty$ and $\alpha >\alpha _u$.  \qed
\end{proof}

\noindent Proofs for higher values of $k$ are 
identical, and will be given in the full version.


\section{The GUC heuristic for random SAT and COL}

The above analysis of the DPLL-UC search tree can be extended to the
GUC heuristic \cite{Fra}, where literals are preferentially  chosen to
satisfy 2-clauses (if any). The outlines of the proofs of Theorems 
2 and 3 are given below; details will be found in the full version. 

{\bf 3-SAT}. 
The main difference with respect to the UC case is that the
two branches issued from the split are not statistically identical. In
fact, the literal $\ell$ chosen by GUC satisfies at least one clause, 
while this clause is reduced to a shorter clause when $\ell$ is set to
False. The cases $C_2\ge 1$ and $C_2=0$ have also to be considered 
separately.
With $f_1,f_2,f_3$ defined in the same way as in the UC case, 
we obtain
\begin{eqnarray}
\label{eqevg}
G(x_1,x_2,x_3&;&T+1\,) = \frac 1{f_1}\; G \big( f_1,f_2,f_3; 
T\,\big) + \bigg( \frac{1+f_1}{f_2} - \frac 1{f_1} \bigg) \; 
G\big( 0, f_2,f_3; T\, \big)  \nonumber \\
&+&  \bigg( \frac{1+f_2}{f_3} - \frac {1+f_1}{f_2} \bigg) \; 
G\big( 0, 0,f_3; T\, \big) - \frac{1+f_2}{f_3} \;  G(0,0,0;T) \ . 
\end{eqnarray}
The asymptotic analysis of $G$ follows the lines of Section 3. 
Choosing $f_2=f_1+f_1^2$ {\em i.e.}
$x_1=(-1+\sqrt{1+4 x_2})/2+O(1/N)$ allows us to cancel the second term on the
r.h.s. of (\ref{eqevg}). Iterating relation (\ref{eqevg}),
we establish the counterpart of Lemma 2 for GUC: the value of $G$ at height
$[tN]$ and argument $x_2,x_3$ is equal to its (known) value at height 0
and argument $y_2,y_3$ times the product of factors $\frac 1{f_1}$, up to an
additive term, $A$, including iterates of the third and fourth terms
on the right hand side of (\ref{eqevg}).
$y_2,y_3$ are the values at 'time' $\tau=0$ of the solutions of
the ordinary differential equations (ODE) $dY_2/d\tau=- 2 m(Y_2)/(1-\tau)$,
$dY_3 /d\tau = - 3 ((1+Y_2)/2-Y_3)/(1-\tau)$
with 'initial' condition $Y_2(t)=x_2$, $Y_3(t)=x_3$
(recall that function $m$ is defined in Theorem 2).  Eliminating 'time'
between $Y_2,Y_3$ leads to the ODE in Theorem 2.
The first term on the r.h.s. in the expression of $\omega
^g$
(\ref{refiuy}) corresponds to  the  logarithm of the
product of factors $\frac 1{f_1}$ between heights $0$ and $T$.
The maximum over $y_2$ in expression (\ref{refiuy}) for $\omega ^g$
is equivalent to the maximum over the reduced height $t$ appearing
in
$\omega _C$ in Theorem 1  (see also Proposition~2).
Finally, choosing $\alpha > \alpha _u^g$ ensures that,
from the one hand,  the additive term $A$ mentioned above
is asymptotically negligible and, from the other hand,
the ratio of the expected sizes of \#DPLL-GUC and DPLL-GUC is
asymptotically equal to unity (see proof of Corollary 1).

{\bf 3-COL}. 
The uniformity expressed by Lemma 1 holds: the 
subgraph resulting from the coloring of $T$ vertices is still
Erd\H{o}s-R\'enyi-like with edge probability $\frac cN$, conditioned to 
the numbers $C_j$ of vertices with $j$ available colors \cite{mol}.
The generating function $G$ of the average
number of residual asignments equals $(x_3)^N$ at height $T=0$ and
obeys the reccurence relation, for $T<N$,
\begin{eqnarray}
\label{eqevgc}
G(x_1,x_2,x_3;T+1\,) &=& \frac 1{f_1}\; G \big( f_1,f_2,f_3; 
T\,\big) + \bigg( \frac{2}{f_2} - \frac 1{f_1} \bigg) \; 
G\big( 0, f_2,f_3; T\, \big) \nonumber \\ &+&  
\bigg( \frac{3}{f_3} - \frac {2}{f_2} \bigg) \; G\big( 0, 0,f_3; T\, \big) 
\end{eqnarray}
with $f_1=(1-\mu)x_1$, $f_2=(1-2\mu)x_2+2\mu x_1$, $f_3=(1-3\mu)x_3+3\mu x_2$,
and $\mu =c/(3N)$.
Choosing $f_1=\frac 12 f_2$ {\em i.e.}
$x_1=\frac 12 x_2+O(1/N)$ allows us to cancel the second term on the
r.h.s. of (\ref{eqevgc}). Iterating relation (\ref{eqevg}), 
we establish the counterpart of Lemma 2 for GUC: the value of $G$ at height 
$[tN]$ and argument $x_2,x_3$ is equal to its (known) value at height 0 
and argument $y_2,y_3$ respectively, 
times the product of factors $\frac 1{f_1}$, up to an
additive term, $A$, including iterates of the last  term in (\ref{eqevgc}).
An explicit calculation  leads to $G(\frac 12
x_2,x_2,x_3;[tN])=e^{N \gamma^h(x_2,x_3,t)+o(N)} +A$ for $x_2,x_3>0$, where 
$\gamma^h(x_2,x_3,t) = \frac c6 t^2 - \frac c3 t +(1-t)\ln (x_2/2)+
\ln[3 + e^{-2ct/3}(2 x_2/x_3-3)]$. As in Proposition 2, we bound
from below (respectively, above) the number of contradiction leaves in
\#DPLL-GUC tree by the exponential of ($N$ times) the value of function
$\gamma ^h$ in $x_2=x_3=1$ at reduced height $t$ (respectively, 
lower than $t$).
The maximum over $t$ in Theorem 3 is equivalent to the maximum over the 
reduced height $t$ appearing in $\omega _C$ in Theorem 1 
(see also Proposition 2). Finally,
we choose $c _u^h$ to make the additive term $A$ negligible.
Following the notations of Corollary 1, we use $L_{sat}\le 3^N$, and
$P_{sat}\le 3^N e^{-Nc/6 + o(N)}$, the expected number of 3-colorings 
for random graphs from $G(N,c/N)$.

\section{Conclusion and perspectives}

We emphasize that the average \#DPLL tree size can be calculated for even more
complex  heuristics e.g. making decisions based on
literal degrees \cite{Kap}. This task requires, in
practice, that one is able: first, to find the correct
conditioning ensuring uniformity along a branch (as 
in the study of DPLL in the absence of backtracking);
secondly, to determine the asymptotic behaviour of the associated 
generating function $G$ from the recurrence relation for $G$.

To some extent,
the present work is an analytical implementation of an 
idea put forward by Knuth thirty years ago \cite{Knu,Coc2}. Knuth
indeed proposed to estimate the average computational effort required 
by a backtracking procedure through successive runs of the 
non--backtracking counterpart, each weighted in an appropriate
way \cite{Knu}. This weight is, in the language of Section II.B, 
simply the probability of a branch (given the heuristic
under consideration) in \#DPLL search tree times
$2^S$ where $S$ is the number of splits \cite{Coc2}.

Since the amount of backtracking seems to have a heavy tail
\cite{Gent,Jia}, 
the expectation is often not a good predictor in practice.
Knowledge of the second moment of the search tree size would be
very precious; its calculation, currently under way, requires us to treat 
the correlations between nodes attached to distinct branches. 
Calculating the second moment is a step towards the distant goal of 
finding the expectation of the logarithm, which
probably requires a deep understanding of correlations as in 
the replica theory of statistical mechanics.
 
Last of all, \#DPLL is a complete procedure for enumeration.
Understanding its average-case operation will, hopefully, provide
us with valuable information  not only on the algorithm itself but also 
on random decision problems e.g. new bounds on the sat/unsat 
or col/uncol thresholds, 
or insights on the statistical properties of solutions.

\noindent
{\bf Acknowledgments:}
The present analysis is the outcome of a work started four years ago
with S. Cocco to which I am deeply indebted \cite{Coc,Coc2}.   
I am grateful to C. Moore for numerous and illuminating
discussions, as well as for a critical reading of the manuscript.
I thank J. Franco for his interest and support, and the referee 
for pointing out Ref. \cite{ben}, the results of which agree with the
$\alpha ^{-1/(k-2)}$ asymptotic scaling of $\omega$ found here.


\end{document}